\documentclass[preprint]{elsarticle}

\usepackage{graphicx}% Include figure files
\usepackage{subfigure}
\usepackage{color}

\journal{arXiv}

\begin{document}

\begin{frontmatter}

%\title{Redfield-theory calculation of spin relaxation and geometric phase
% for a rectangular prism cell with arbitrary, time-independent field gradient.}
\title{Spin relaxation and linear-in-electric-field frequency shift
 in an arbitrary, time-independent magnetic field}

\author{Steven M. Clayton\footnote{\emph{Present address:} Los Alamos National Laboratory, PO Box 1663, MS H846, Los Alamos, NM, 87545}}
\ead{sclayton@lanl.gov}
\address{Department of Physics, University of Illinois, Urbana, IL, 61820}


\begin{abstract}
A method is presented to calculate the spin relaxation
times $T_1$, $T_2$ due to a non-uniform magnetic field,
and the linear-in-electric-field precession frequency shift $\delta\omega_E$
when an electric field is present,
in the diffusion approximation for spins confined to a rectangular cell.
It is found that the rectangular cell
geometry admits of a general result
for $T_1$, $T_2$, and $\delta\omega_E$
in terms of the spatial cosine-transform components of the magnetic field.
\end{abstract}

\begin{keyword}


\end{keyword}

\end{frontmatter}

%%%%%%%%%%%%%%%%%%%%%%%%%%%%%%%%%%%%%%%%%%%%%%%%%
\section{Introduction}

An experiment to measure the neutron electric dipole moment (nEDM),
to be installed at the FnPB beamline at Oak Ridge National Laboratory,
will utilize a helium-3 comagnetometer in the central, superfluid-helium-filled
measurement cell~\cite{nedm_phys_report}\cite{nedm_proposal}.
The helium-3 polarization must remain high over the entire
measurement period, $\sim$1000~seconds, as the helium atoms
precess in the holding field and diffuse within a rectangular cell.
Also, as there is a strong electric field $\vec{E}$ applied across the cell,
a subtle effect,
in which the interplay of the motional $\vec{v}\times\vec{E}$ field with gradients in
the static magnetic field cause the precession frequency
to shift linearly with $\vec{E}$~\cite{vxE_ref},
must be well-understood or shown to be negligible.
Design optimization of the experimental apparatus includes calculating
the helium-3 spin relaxation times $T_1$, $T_2$, and linear-in-electric-field
frequency shift $\delta\omega_E$ due to given magnetic field non-uniformities.

In this article, a method is shown to calculate these quantities in the
diffusion approximation in a rectangular cell and for an arbitrary magnetic field.
The starting point for the relaxation times is the
Redfield theory of spin relaxation~\cite{redfield}.
In second order perturbation theory these
can be written, for a holding field in the $z$ direction, as~\cite{slichter}
\begin{equation}\label{eq:T1}
\frac{1}{T_1} = \gamma^2 \left(k_{xx}(\omega_0) + k_{yy}(\omega_0) \right),
\end{equation}
\begin{equation}\label{eq:T2}
\frac{1}{T_2} = \frac{1}{2 T_1} + \gamma^2 k_{zz}(0),
\end{equation}
where the spectral density is given in terms of magnetic field perturbations
$h_i(t)$,
\begin{equation}\label{eq:kqq_def}
k_{ij}(\omega) = \frac{1}{2}\int_{-\infty}^{\infty} \langle h_i(t) h_{j}(t+\tau) \rangle \cos \omega\tau d\tau
\end{equation}
Here, the total field in each direction $i$ is
$H_i(t) = \langle H_i(t) \rangle + h_i(t)$, such that
average perturbation $\langle h_i(t) \rangle = 0$,
and $\omega_0 = \gamma \langle H_z(t) \rangle$ is the average
spin precession frequency.

McGregor~\cite{mcgregor90} calculated the ensemble average correlation of the field
perturbations seen by a diffusing particle in the case of a time-independent,
uniform gradient of $H_z$ in the $x$-direction,
\begin{equation}
\langle h_z(t) h_z(t+\tau) \rangle = \left(\frac{\partial H_z}{\partial x}\right)^2
 \langle x(t) x(t+\tau) \rangle,
\end{equation}
resulting in an analytic expression for $T_2$ in a rectangular cell,
\begin{equation}
\frac{1}{T_2} = \frac{1}{2 T_1} + \frac{\gamma^2 L^4}{120 D}
 \left[\frac{\partial H_z}{\partial x}\right]^2.
\end{equation}

We relax the requirement of uniform gradient and find,
in the case of a rectangular prism cell, that $T_1$ and $T_2$ can
be written in terms of the components of the 3D cosine transform
of $h_q(\vec{r})$ over the cell volume.
The same technique is applied to
dressed spins~\cite{spin_dressing_ref}, with
uniform holding field and non-uniform dressing field, by mapping non-uniformities
in the dressing field to equivalent non-uniformities in the holding field.
In Section~\ref{sec:vxE}, a variation of the technique is used
for the linear-in-electric-field frequency shift.
Finally, as an example, in Section~\ref{sec:example}
the relaxation times and linear-in-electric-field
frequency shift due to a superconducting rod near the cell
are computed.

%%%%%%%%%%%%%%%%%%%%%%%%%%%%%%%%%%%%%%%%%%%%%%%%%%%%
\section{Correlation functions in the diffusion limit}\label{sec:corr}

The correlation function of $h_i$ can be expressed as integrals
over the cell volume weighted by the probability density $p(\vec{r}_0,t)$
that the particle is at $\vec{r}_0$ at the initial time $t_0$,
and the joint probability density $p(\vec{r},t|\vec{r}_0,t_0)$
that a particle at $\vec{r}_0$ at time $t_0$ will be at $\vec{r}$ at time
$t$.
Thus,~\cite{mcgregor90}
\begin{equation}\label{eq:hqq_tau}
\langle h_i(t_0) h_i(t_0+\tau) \rangle = \int_V d\vec{r}_0 h_i(\vec{r}_0) p(\vec{r}_0,t_0)
 \int_V d\vec{r} h_i(\vec{r}) p(\vec{r},t_0+\tau | \vec{r}_0, t_0).
\end{equation}
The particle density will be taken as uniform in the cell, $p(\vec{r},t) = 1/V$.
The joint probability is the solution to the diffusion equation,
\begin{equation}
\frac{\partial}{\partial t} p(\vec{r},t | \vec{r}_0, t_0)
  = D \nabla^2 p(\vec{r},t | \vec{r}_0, t_0),
\end{equation}
subject to reflecting boundary conditions at the walls,
\begin{equation}
\nabla p(\vec{r}_S,t | \vec{r}_0, t_0) \cdot \hat{n} = 0,
\end{equation}
where $\vec{r}_S \in S$ and $\hat{n}$ is normal to the wall.
In a rotated coordinate system $(x',y',z')$ aligned with the cell walls,
the solution for a $L_x \times L_y \times L_z$ box with walls
at $x'=\pm L_x/2$, $y'=\pm L_y/2$ and $z'=\pm L_z/2$ is
\begin{equation}\label{eq:prtr0t0}
 p(\vec{r}',t|\vec{r}'_0, t_0) = p(x',t|x'_0,t_0;L_x) p(y',t|y'_0,t_0;L_y) p(z',t|z'_0,t_0;L_z),
\end{equation}
%%%%%%%%%%%
with the 1D solution dependent on the time difference $\tau = t - t_0$,~\cite{mcgregor90}
\begin{eqnarray}
 p(x',t | x'_0, t_0;L_x) &=& \frac{1}{L_x}
  + \frac{2}{L_x} \sum_{n=1,3,...}^{\infty} e^{-n^2\pi^2 D\tau/L_x^2}
  \sin\frac{n\pi x'}{L_x} \sin\frac{n\pi x'_0}{L_x} \nonumber \\
  & & + \frac{2}{L_x} \sum_{n=2,4,...}^{\infty} e^{-n^2\pi^2 D\tau/L_x^2}
  \cos\frac{n\pi x'}{L_x} \cos\frac{n\pi x'_0}{L_x}.
\end{eqnarray}
It will be convenient to recognize the following:
\begin{equation}\label{eq:prtr0t0_shifted}
 p(x'-L_x/2,t | x'_0-L_x/2, t_0;L_x) = \frac{1}{L_x}
  + \frac{2}{L_x} \sum_{n=1,2,3,...}^{\infty} e^{-n^2\pi^2 D\tau/L_x^2}
  \cos\frac{n\pi x'}{L_x} \cos\frac{n\pi x'_0}{L_x}.
\end{equation}

Putting Eq.~\ref{eq:prtr0t0} and $p(\vec{r}'_0,t_0) = 1/V$ into Eq.~\ref{eq:hqq_tau},
changing the limits of integration to $0 \leq q'_i \leq L_i$ for each
dimension $q'_i$ and using Eq.~\ref{eq:prtr0t0_shifted}, we have
\begin{eqnarray}\label{eq:hqq_tau2}
g_{ii}(\tau) &=& \langle h_i(t_0) h_i(t_0 + \tau) \rangle \nonumber \\
  &=& \sum_{n_x,n_y,n_z=0}^{\infty}
 e^{-\pi^2 D (n_x^2/L_x^2 + n_y^2/L_y^2 + n_z^2/L_z^2)\tau} \nonumber \\
 & & \times\frac{1}{V}\int_{V'} dx'_0\, dy'_0\, dz'_0\, h_i(x'_0-L_x/2,y'_0-L_y/2,z'_0-L_z/2) \nonumber \\
 & & C_{n_x}C_{n_y}C_{n_z}\cos\frac{n_x\pi x'_0}{L_x}\cos\frac{n_y\pi y'_0}{L_y}
  \cos\frac{n_z\pi z'_0}{L_z} \nonumber \\
 & & \times\frac{1}{V}\int_{V'} dx'\, dy'\, dz'\, h_i(x'-L_x/2,y'-L_y/2,z'-L_z/2) \nonumber \\
 & & C_{n_x}C_{n_y}C_{n_z}\cos\frac{n_x\pi x'}{L_x}\cos\frac{n_y\pi y'}{L_y}
  \cos\frac{n_z\pi z'}{L_z}
\end{eqnarray}
%%%%%%%%%%%%%%%%%%%%%
in which the factor $C_n$ has been introduced,
\begin{equation}
 C_n = \left\{
  \begin{array}{l l}
    1 & \quad \mbox{if $n = 0$}\\
    \sqrt{2} & \quad \mbox{otherwise.} \end{array} \right.
\end{equation}
We identify the 3D cosine transform of $h_i(\vec{r}')$ within Eq.~\ref{eq:hqq_tau2},
\begin{eqnarray}\label{eq:An}
 \mathcal{A}^{\vec{n}}\{ h_i \} &\equiv& \frac{1}{V}\int_{V'} dx'\, dy'\, dz'\, h_i(x'-L_x/2,y'-L_y/2,z'-L_z/2) \nonumber \\
 & & \times \cos\frac{n_x\pi x'}{L_x}\cos\frac{n_y\pi y'}{L_y}
  \cos\frac{n_z\pi z'}{L_z},
\end{eqnarray}
(where the integral is over the range $x' \in [0,L_x]$, $y' \in [0,L_y]$, $z' \in [0,L_z]$),
giving finally
\begin{equation}\label{eq:gqq}
g_{ii}(\tau) = \sum_{n_x,n_y,n_z=0}^{\infty}
 e^{-\tau/\tau_c^{\vec{n}}}
 C_{n_x}^2 C_{n_y}^2 C_{n_z}^2 \left( \mathcal{A}^{\vec{n}}\{ h_i \} \right)^2,
\end{equation}
with the characteristic time $\tau_c$ for a given spatial mode defined by
\begin{equation}
\frac{1}{\tau_c^{\vec{n}}} = \pi^2 D \left( \frac{n_x^2}{L_x^2} + \frac{n_y^2}{L_y^2}
 + \frac{n_z^2}{L_z^2} \right)
\end{equation}
Putting this expression into Eq.~\ref{eq:kqq_def} and performing the
integral over $\tau$ gives
\begin{equation}\label{eq:kqq}
k_{ii}(\omega) = \sum_{n_x,n_y,n_z = 0}^{\infty}
 C_{n_x}^2 C_{n_y}^2 C_{n_z}^2
 \frac{\tau_c^{\vec{n}}}{1 + \omega^2 (\tau_c^{\vec{n}})^2} \left( \mathcal{A}^{\vec{n}}\{ h_i \} \right)^2,
\end{equation}
Substitution into Eqs.~\ref{eq:T1} and~\ref{eq:T2} results in complete
expressions for the longitudinal and transverse relaxation times.

% The result also applies to an
% average field $\langle \vec{H}(\vec{r}) \rangle$
% in an arbitrary direction $\hat{z}'$, rather than along $\hat{z}$.
% We note that the subscripts in Eqs.~\ref{eq:T1} and~\ref{eq:T2}
% refer to a coordinate system defined with respect to the holdling field.
% The spatial cosine transforms, in the (unprimed) coordiate system aligned with the
% rectangular cell, operate on the 
% 
For arbitrary fields including field maps,
the cosine-transform amplitudes $\mathcal{A}$ can be numerically calculated by
fast discrete cosine transform (DCT) over $h_q(\vec{r})$ sampled at
a sufficient number of points throughout the cell volume.
In this case the summation over $n_q$ is truncated accordingly, and accuracy is checked
by increasing the number of sample points and comparing results.

%%%%%%%%%%%%%%%%%%%%%%%%%%%%%%%%%%%%%%%%%%%%%%%%%%%%%%%%%
\subsection{Extension to dressed spins with non-uniform dressing field}\label{sec:dressed}

An RF magnetic field with amplitude $B_1$ applied transverse to the holding field $B_0$
modifies the effective precession frequency of a particle.  In terms of dimensionless
``dressing parameters''~\cite{chu_dressed_spin}
%\begin{eqnarray}
%X &=& \frac{\gamma B_1}{\omega_{\rm RF}}, \\
%Y &=& \frac{\gamma B_0}{\omega_{\rm RF}},
%\end{eqnarray}
\begin{equation}
X = \frac{\gamma B_1}{\omega_{\rm RF}},\ Y = \frac{\gamma B_0}{\omega_{\rm RF}},
\end{equation}
in the limit $Y \ll 1$ the effective gyromagnetic ratio becomes~\cite{spin_dressing_ref}
\begin{equation}
\gamma_{\rm eff} = \gamma J_0(X).
\end{equation}
Thus for a dressing field with spatially varying amplitude
$B_1(\vec{r}) = \langle B_1 \rangle + \delta B_1(\vec{r})$,
the equivalent variation $\delta\tilde{B}_0$
in the holding field $B_0$ is given by~\cite{nedm_phys_report}
\begin{equation}
\gamma_{\rm eff} \delta \tilde{B}_0 \leftrightarrow \delta\gamma_{\rm eff} B_0
           = B_0\frac{\partial\gamma_{\rm eff}}{\partial X} \delta X
           = B_0\gamma J_1(X) \delta X,
% \delta \tilde{B}_0 &\leftrightarrow& B_0\frac{J_1(X)}{J_0(X)} \frac{\gamma \delta B_1}{\omega_{\rm RF}}
%           = Y\frac{J_1(X)}{J_0(X)} \delta B_1.
\end{equation}
The expressions from the previous sections can be used to calculate $T_2$
for the dressed spin with a non-uniform dressing field by setting
\begin{eqnarray}
\gamma &\rightarrow& \gamma_{\rm eff},\\
\omega_0 &\rightarrow& \gamma_{\rm eff} B_0,\\
h_z(\vec{r}) &\rightarrow& Y\frac{J_1(X)}{J_0(X)} \delta B_1(\vec{r}) + \delta B_0(\vec{r}),
\end{eqnarray}
where the variation $\delta B_0$ in the holding field itself has been included in $h_z$.

%%%%%%%%%%%%%%%%%%%%%%%%%%%%%%%%%%%%%%%%%%%%%%%%%%%%%%%%%
\section{Linear electric field frequency shift}\label{sec:vxE}

A spin moving though an electric field experiences a motional magnetic field
that may, in conjunction with gradients of the magnetic field, produce a shift
in the precession frequency dependent on the electric field direction and magnitude.~\cite{vxE_ref}
Of particular concern in searches for electric dipole moments are effects that
are linearly proportional to the electric field $E$.  These may mimic
effects expected for an electric dipole moment, thereby creating a ``false EDM.''

As shown by Lamoreaux and Golub~\cite{lamgol05_vxE},
the linear-in-electric-field frequency shift for spins in a confined volume
is given by the expression
\begin{equation}
\delta\omega_{E} = -\frac{1}{2} \int_0^t d\tau\,\cos \omega_0\tau \{
 \langle \omega_x(t)\omega_y(t-\tau) \rangle - \langle \omega_x(t-\tau)\omega_y(t) \rangle \},
\end{equation}
where the perturbations can be written more generally as
\begin{equation}
 \omega_x(t) = \gamma h_x(t) + \gamma\frac{E}{c} v_y(t),
 \ \ \omega_y(t) = \gamma h_y(t) - \gamma\frac{E}{c} v_x(t).
\end{equation}
Expanding the expression for $\delta\omega_{E}$ and keeping only terms linear in $E$
results in
\begin{eqnarray}\label{eq:deltaomega}
\delta\omega_{E} &=& \frac{\gamma^2 E}{2 c} \int_0^t d\tau\,\cos \omega_0\tau \{
 \langle h_y(t) v_y(t-\tau) \rangle
 + \langle h_x(t) v_x(t-\tau) \rangle \nonumber\\
 & & - \langle h_y(t-\tau) v_y(t) \rangle
 - \langle h_x(t-\tau) v_x(t) \rangle
 \}.
\end{eqnarray}
%Golub has suggested~\cite{gol08_vxE_suggestion} that the cosine-transform
The cosine-transform
method developed in the present work can be used to compute
Eq.~\ref{eq:deltaomega} in the diffusion limit for the nEDM cell geometry.
While the expressions,
\begin{equation}\label{eq:yFTC}
 y(t) = y_0 + \int_0^t v_y(t') dt',\ \ y(t-\tau) = y_0 + \int_0^{t-\tau} v_y(t') dt',
\end{equation}
were used in Ref.~\cite{lamgol05_vxE}
to eliminate $y$ in favor of an expression with $v_y$,
here we remove the velocity components from the correlation
functions and use instead the Fundmental Theorem of Calculus and the above
expression for $y(t-\tau)$.
The correlation functions can then be written
\begin{equation}
\langle h_y(t) v_y(t-\tau) \rangle
 = -\frac{\partial}{\partial\tau} \langle h_y(t) y(t-\tau) \rangle,
\end{equation}
\begin{equation}
\langle h_y(t - \tau) v_y(t) \rangle
 = \langle h_y(t) v_y(t + \tau) \rangle
 = \frac{\partial}{\partial\tau} \langle h_y(t) y(t + \tau) \rangle.
\end{equation}
In the latter expression, averages are assumed to be independent
of the overall time offset $t$, as appropriate for a stationary problem.

The expressions in Section~\ref{sec:corr} are modified to give the correlation
function in the diffusion limit in terms of the cosine-transform components
of $h_y(\vec{r})$ and $y(\vec{r})$,
\begin{equation}
\langle h_y(t) y(t-\tau) \rangle = \sum_{n_x,n_y,n_z} e^{-\tau/\tau_c^{\vec{n}}}
 C_{n_x}^2 C_{n_y}^2 C_{n_z}^2
\mathcal{A}^{\vec{n}}\{ h_y \}\, \mathcal{A}^{\vec{n}}\{ y \}.
\end{equation}

\begin{equation}
\langle h_y(t) v_y(t-\tau) \rangle = \sum_{n_x,n_y,n_z} \frac{1}{\tau_c^{\vec{n}}}
 e^{-\tau/\tau_c^{\vec{n}}}
 C_{n_x}^2 C_{n_y}^2 C_{n_z}^2
 \mathcal{A}^{\vec{n}}\{ h_y \}\, \mathcal{A}^{\vec{n}}\{ y \}
\end{equation}

Performing the integral in Eq.~\ref{eq:deltaomega}, we have
\begin{equation}
\int_0^{\infty} d\tau\,\cos\omega_0\tau \langle h_y(t) v_y(t-\tau) \rangle
 = \sum_{n_x,n_y,n_z} \frac{1}{1 + (\omega_0 \tau_c^{\vec{n}})^2}
 C_{n_x}^2 C_{n_y}^2 C_{n_z}^2
 \mathcal{A}^{\vec{n}}\{ h_y \}\, \mathcal{A}^{\vec{n}}\{ y \},
\end{equation}
leading to an expression for the frequency shift,
\begin{equation}
\delta\omega_{E} = \frac{\gamma^2 E}{c}
 \sum_{n_x,n_y,n_z} \frac{1}{1 + (\omega_0 \tau_c^{\vec{n}})^2}
 C_{n_x}^2 C_{n_y}^2 C_{n_z}^2
 \left[ \mathcal{A}^{\vec{n}}\{ h_y \}\, \mathcal{A}^{\vec{n}}\{ y \}
 + \mathcal{A}^{\vec{n}}\{ h_x \}\, \mathcal{A}^{\vec{n}}\{ x \} \right].
\end{equation}
The summation can be reduced by computing the cosine transform components
of $x(\vec{r})$ and $y(\vec{r})$ analytically,
\begin{equation}
\mathcal{A}^{\vec{n}}\{ x \} = \left\{ \begin{array}{ll}
  -\frac{2 L_x}{n_x^2 \pi^2} & \mbox{if $n_x = 1,3,...; n_y=n_z=0$} \\
  0   & \mbox{otherwise.}
  \end{array} \right.
\end{equation}
The result for the frequency shift in the diffusion approximation is
\begin{equation}\label{eq:dw_result}
\delta\omega_{E} = \frac{4 \gamma^2 E}{c} \left[
 \sum_{n_y=1,3,...} \frac{L_y}{n_y^2 \pi^2}
  \frac{\mathcal{A}^{(0,n_y,0)}\{ h_y \}}{1 + (\omega_0 \tau_c^{(0,n_y,0)})^2}
 + \sum_{n_x=1,3,...} \frac{L_x}{n_x^2 \pi^2}
  \frac{\mathcal{A}^{(n_x,0,0)}\{ h_x \}}{1 + (\omega_0 \tau_c^{(n_x,0,0)})^2}
 \right].
\end{equation}

%%%%%%%%%%%%%%%%%%%%%%%%%%%%%%%%%%%%%%%%%%%%%%%%%%%%%%%%%
\section{Example application employing the discrete cosine transform}\label{sec:example}

The bulk of the computational effort required for practical application of the present
technique is in finding the cosine transform amplitudes $\mathcal{A}^{\vec{n}}$
of the field non-uniformities.  The form of Eq.~\ref{eq:An} is amenable to 
numerical computation with Fast Fourier Transform software libraries.
The example below uses the multidimensional discrete cosine transform (DCT)
feature of the freely-available
software library FFTW3~\cite{fftw3_ref}.
Input to the DCT for each field component
is an array of field perturbations $h_q(\vec{r})$ sampled over the cell
volume at $N_x N_y N_z$ grid points, and the output is an array of
DCT amplitudes which, after scaling by $1/(8 N_x N_y N_z)$, correspond
to the desired amplitudes $\mathcal{A}^{\vec{n}}$.
The summations in Eqs.~\ref{eq:kqq} and~\ref{eq:dw_result}
are truncated according to the amplitudes available from the DCT.
Accuracy of the result can be checked by increasing the number of sample points
and repeating the computation.

\begin{figure}
\subfigure{\includegraphics[width=10cm]{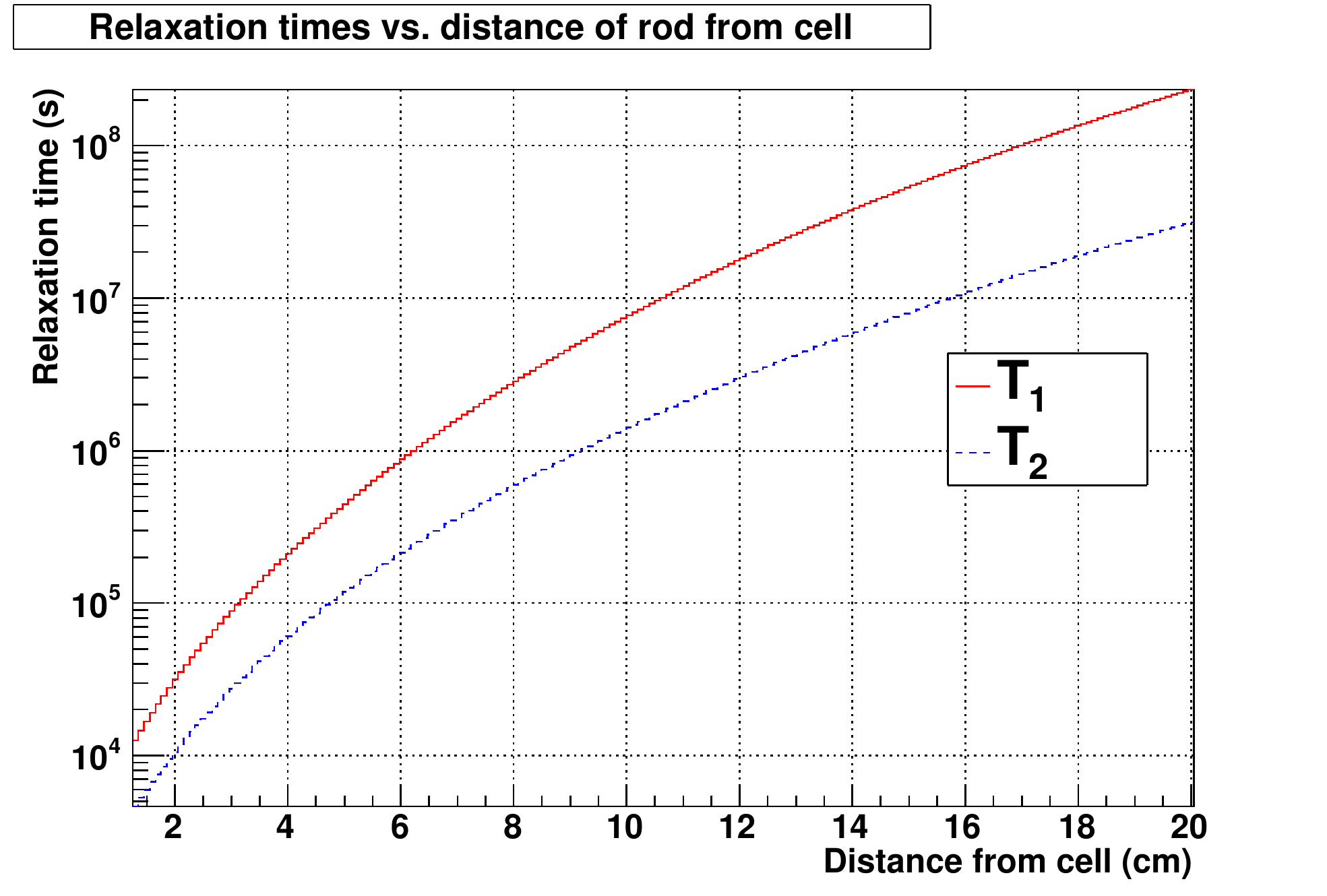}\label{fig:T1T2}}
\subfigure{\includegraphics[width=10cm]{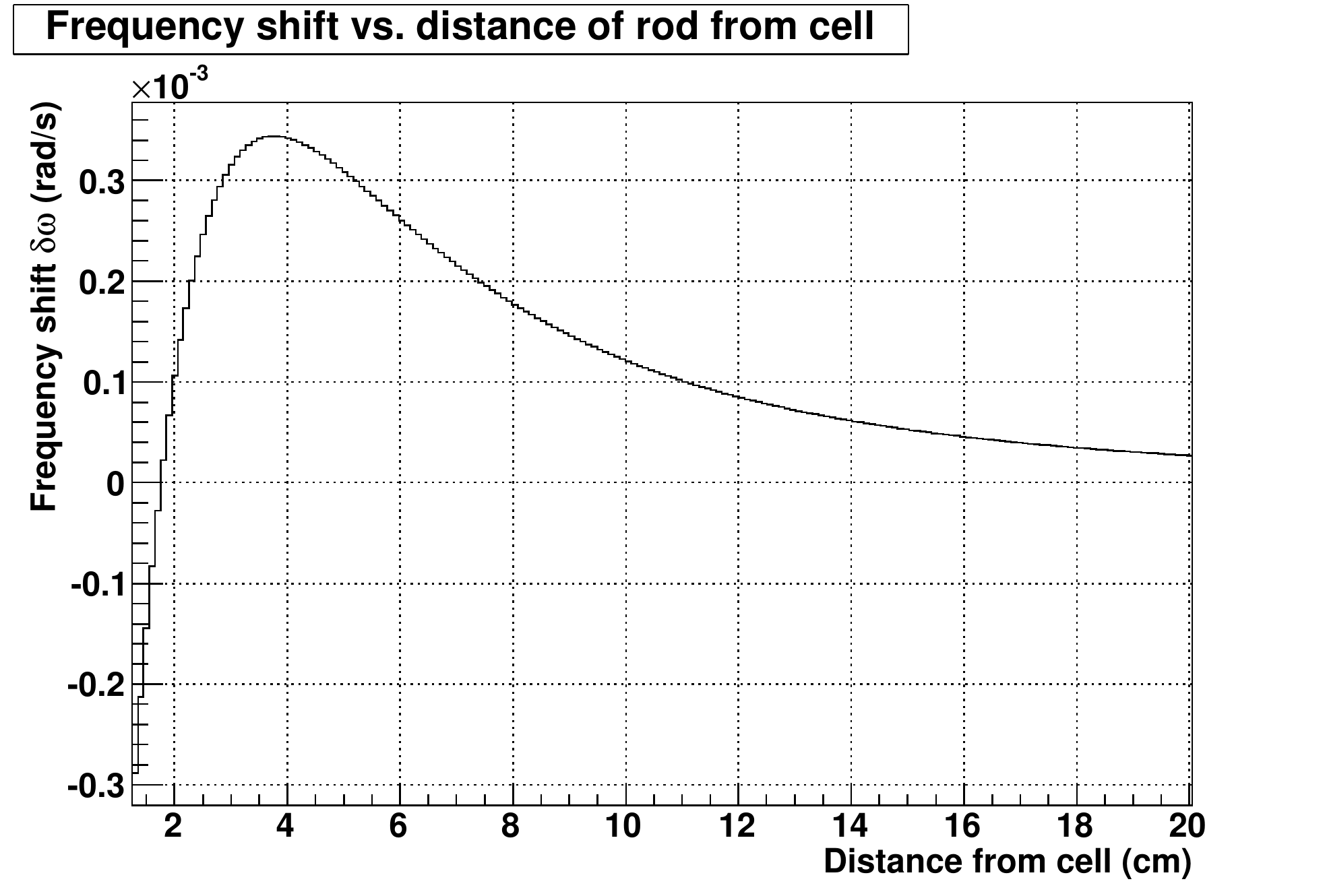}\label{fig:freq_shift}}
\caption{Calculated \subref{fig:T1T2} relaxation times $T1$, $T2$ and
\subref{fig:freq_shift} frequency shift $\delta\omega_E$ due to the magnetic
field distortion from a 3-mm-diameter,
infinitely-long superconducting rod parallel to the $z$-axis and placed near the cell.
The parameters are $\gamma = -20393.963\ {\rm rad\,G^{-1} s^{-1}}$
(gyromagnetic ratio for $^3$He), $D = 428\ \rm{cm^2/s}$
(corresponding to a temperature of 450~mK~\cite{lam02_diffusion_coeff}),
 $B_0 = 10$~mG (applied parallel to the $x$-axis), $E = 100\ \rm{kV/cm}$ (parallel to $B_0$),
 cell dimensions $L_x = 7.8$~cm, $L_y = 10$~cm, $L_z = 40$~cm.
}\label{fig:example}
\end{figure}

%%%%%%%%%%%%%%%%%%%%%%%%%
We calculate the effect of a superconducting rod placed near the cell
in an otherwise uniform holding field.
For an infinite-length rod along the $z$ axis and through the origin in a magnetic field
$B_0$ applied along the $x$ axis, the net field around the rod is~\cite{poole}
\begin{eqnarray}
B_\rho &=& (1 - \frac{a^2}{\rho^2}) B_0 \cos{\phi},\\
B_\phi &=& (1 + \frac{a^2}{\rho^2}) B_0 \sin{\phi},
% B_x &=& B_\rho \cos{\phi} + B_\phi \sin{\phi},\\
% B_y &=& -B_\rho \sin{\phi} + B_\phi \cos{\phi},
\end{eqnarray}
where $a$ is the rod radius and $(\rho, \phi)$ are polar coordinates in the $xy$ plane.
While the cosine transforms of $B_x$, $B_y$ derived from these equations
(appropriately translated to the desired location of the superconducting rod)
could perhaps be calculated analytically,
here the equations are used to generate a field map that is subsequently run through the
machinery to produce values for $T_1$, $T_2$, and $\delta\omega_E$.
Results are shown in Figure~\ref{fig:example}, with physical parameters
given in the caption.
% The relaxation times are found to be sufficiently
% long, not surprising since we are in the extreme motional narrowing
% regime~\cite{motional_narrowing_ref}.
% However, even with the rod 20~cm away from the cell,
% the linear (in $E$) frequency shift turns out to be larger than acceptable
% for the nEDM experiment.

% Critical dressing for the neutron and $^3$He is at $X_n \approx 1.189$, where
% the subscript $n$ refers to $X$ in terms of the neutron gyromagnetic ratio $\gamma_n$,
% or $X_3 \approx 1.323$ where the subscript refers to $^3$He.
% The parameters in Riccardo's simulation were $B_0 = 10\ {\rm mG}$ and
% $\omega_{\rm RF} = 2\times 10^4\ {\rm rad/s}$, for a critical dressing field amplitude
% of $B_1 \sim 1.3\ {\rm G}$.  Reading off the plot in Riccardo's slide~2, the (quadratic)
% dressing field deviation is $\Delta B_1/B_1 \approx 4.8\times 10^{-3}$ at $z = \pm 20$~cm.
% We see that the machinery of the Redfield/generalized McGregor calculation
% reproduces the simulation result, as does Riccardo's modified McGregor formula
% as noted in the teleconference.  However, in the present method, $T_2$ can be
% calculated for an arbitrary field variation with no requirement on the functional
% form.  Note: in the Redfield/generalized McGregor theory,
% in the limit $T_1 \gg T_2$, $T_2$ is strictly proportional to the diffusion
% coefficient $D$.

\section{Conclusion}

A method to calculate spin relaxation times and the linear-in-$E$ frequency shift
in the diffusion approximation was presented.
The technique is based on the observation that, for a particle diffusing in
a rectangular cell, the correlation function of a position-dependent
field is the weighted sum over the squared cosine-transform components of the field
(see Eq.~\ref{eq:gqq}).\footnote{During preparation of this manuscript,
independent work was published based on essentially
the same observation.
These authors point out that their result may be used as a probe for possible
unknown spin interactions and applied it to improve the limits on axion-like
interactions with the cell walls~\cite{petukhov10}.}

As the formulation is intended for practical computation
 in a rectagular cell, the result is for the complete 3-dimensional geometry.
It could also be applied to a cylindrical cell if the field variation
 in coordinates transverse to the cylinder axis can be neglected.
Actual computation is done using fast discrete cosine transforms of the
 field components.
This method could be used for example in magnet coil design optimization:
given a field map, $T_1$, $T_2$, and $\delta\omega_{E}$
can be quickly evalulated and combined into a figure of merit.
% Alternatives for an arbitrary field are to use computationally intense
% Monte Carlo or to characterize the field
% by an average gradient and apply e.g. results from Ref.~\cite{mcgregor90}.

\section{Acknowledgements}

This work was supported by NSF grant number NSF06-01067.
The author thanks R.~Golub for suggesting the application to the linear-in-electric-field
frequency shift.

\bibliographystyle{Bibstyles/elsarticle-num-names}
\bibliography{clayton_redfield_calc_JMR.bib}

\end{document}